%% file: msbrm.tex
\input brmacro.tex  
\input psfig.sty

\documentstyle[11pt,aaspp]{article}
\slugcomment{To be submitted for publication in the ApJ}

\begin{document}

\title{An Alignment Effect in FR I Radio Galaxies: U-band Polarimetry 
of the Abell 2597 Cluster Central Galaxy}

\author{Brian R. McNamara\altaffilmark{1}}
\affil{Harvard-Smithsonian Center for Astrophysics\\
60 Garden St.\\
Cambridge, MA 02138}

\author{Buell T. Jannuzi}
\affil{National Optical Astronomy Observatories\\
P.O. Box 26732\\
Tucson, AZ 85726-6732}

\author{Craig L. Sarazin}
\affil{Astronomy Department\\
University of Virginia\\
P.O. Box 3818\\
Charlottesville, VA.  22903-0818}

\author{Richard Elston}
\affil{University of Florida\\
Department of Astronomy\\
211 Space Sciences Bldg.\\
Gainesville, FL 32611-2055}

\author{Michael Wise}
\affil{MIT\\
Center for Space Research\\
MS 37-667\\
Cambridge, MA.  02139-4307}

% Notice that each of these authors has alternate affiliations, which
% are identified by the \altaffilmark after each name.  The actual alternate
% affiliation information is typeset in footnotes at the bottom of the
% first page, and the text itself is specified in \altaffiltext commands.
% There is a separate \altaffiltext for each alternate affiliation
% indicated above.

\altaffiltext{1}{Visiting Astronomer, Kitt Peak National Observatory. 
KPNO is operated by AURA, Inc.\ under contract to the National Science
Foundation.} 
%\altaffiltext{2}{Society of Fellows, Harvard University.} 

% The abstract environment prints out the receipt and acceptance dates
% if they are relevant for the journal style.  For the aasms style, they
% will print out as horizontal rules for the editorial staff to type
% on, so long as the author does not include \received and \accepted
% commands.  This should not be done, since \received and \accepted dates
% are not known to the author.

\newpage 

\begin{abstract}
We have obtained U-band polarimetry 
of the spatially extended, blue optical continuum associated with the 
FR I radio source PKS 2322$-$122. PKS 2322$-$122 is located in the  
Abell 2597 cluster central galaxy.  We find a three sigma upper limit 
to the degree of polarization of the optical continuum
of less than $6\%$.  The accuracy of the 
measurement is limited primarily by our ability to measure the amount
of diluting galactic starlight.
This limit is inconsistent with the blue continuum 
being primarily scattered light or synchrotron radiation.  We 
can therefore exclude models which attribute the blue continuum to
scattered light from an active nucleus that is hidden from
direct view.  Our result does not support the unification paradigm
for BL Lac objects and FR I radio sources.

Essentially all of the data pertaining to the blue continuum 
along the radio source--the ``blue lobes''--indicate that they
are regions
of recent star formation.  The spatial coincidence between 
the blue lobes and disturbances in the radio source suggests that
star formation may have been induced by an interaction
between the radio source and the cool ($<10^4$ K), surrounding gas.
This result, in addition to the results of a similar study of 
the A1795 cluster central galaxy, shows that
under the appropriate conditions 
FR I radio sources may be capable of inducing significant episodes of 
star formation in elliptical galaxies.  

We compare the restframe U-band polarized
luminosities and 1.4 GHz radio powers of A2597 and A1795
to those for several high redshift radio
galaxies exhibiting the alignment effect.
We find that if the polarized luminosities of radio galaxies 
scale in proportion to their radio luminosities, we would not have
detected a polarized signal in either A2597 or A1795.  
We suggest that the fundamental property distinguishing 
some powerful, high redshift radio galaxies exhibiting the
alignment effect, and the lower-power, FR I radio galaxies  
in cooling flows that exhibit the blue lobe phenomenon, is the strength
of the AGN.  While both FR I and FR II radio sources seem to be capable of 
triggering star formation, FR I radio sources seem to be incapable of
producing a large enough polarized luminosity to contribute significantly 
to the aligned continuum radiation.  
\end{abstract}

% That's it for the front matter.  On to the main body of the paper.
% We'll only put in tutorial remarks at the beginning of each section
% so you can see entire sections together.
%
% In the first two sections, you should notice the use of the LaTeX \cite
% command to identify citations.  The citations are tied to the
% reference list via symbolic tags.  We have chosen the first three
% characters of the first author's name plus the last two numeral of the
% year of publication.  The corresponding reference has a \bibitem
% command in the reference list below.
%
% Please go to the LaTeX manual for a complete description of the
% \cite-\bibitem mechanism.

\section{Introduction}

The well known tendency for the blue optical continuum in
distant, luminous radio galaxies to be aligned with their
radio sources represents a potentially important evolutionary 
phase of radio galaxies.  The blue 
optical continuum associated with the Fanaroff-Riley class two (FR II)
radio sources in powerful radio galaxies often extends several
tens of kiloparsecs into their halos along the axis of the radio
source.  The extended radio 
and optical continuum is usually associated with bright nebular emission.
This so-called  ``alignment effect'' becomes  
increasingly prominent in powerful radio galaxies
at redshifts $z \gae 0.6$, but is rarely strong  
in radio galaxies at redshifts less than $z=0.1$ (see McCarthy 1993
for a review).

Distant, FR II radio sources are frequently bent and distorted in the vicinity
of strong nebular emission.  The nebular gas velocity fields are usually
turbulent and shearing, with characteristic velocities of several hundreds of 
kilometers per second.  Such properties seem to indicate that
strong interactions (e.g. momentum exchange, photon heating and
ionization) are occurring between the radio source and the nebular gas.
Remarkably, processes associated with the radio source,
which originates in a powerful nuclear engine, can affect the 
photometric properties of the host galaxy to large galactic distances.

Several emission mechanisms have been proposed to explain
the extended, blue optical continuum
including star formation, scattered light from an obliquely-directed
active nucleus, synchrotron radiation, nebular
continuum, and inverse Compton radiation (di Serego Alighieri \etal\
1989; Fabian 1989; De Young 1989; Rees 1989;
Begelman \& Cioffi 1989; Daly 1990, 1992; Dickson \etal\ 1995).
Although each of these emission mechanisms is interesting in its
own right, collectively they imply
essentially two important consequences for the evolution
of radio galaxies.  Star formation
at the implied levels of several tens to over a thousand solar masses
per year (e.g. 4C41.17, Dey \etal\ 1997) would
have a significant impact on the development of the stellar composition
and structure of the host galaxy.  The remaining emission mechanisms in general
imply the presence of a powerful central engine, a dense 
interstellar medium, and strong magnetic fields, but may be otherwise
inconsequential to the development of the stellar composition
and structure of the galaxy.  However, the alignment effect cannot be
reduced to a unique mechanism, as one or more of these emission mechanisms
may contribute significantly to the effect in a particular galaxy 
and among galaxies.  

It has been known for some time that central dominant cluster 
galaxies (CDGs) selected on the basis of high surface brightness X-ray
emission tend to have unusually blue cores accompanied by
bright, spatially extended nebular emission, and 
Fanaroff-Riley class 1 (FR I) radio sources 
(Baum 1992; Cardiel, Gorgas, \& Aragon-Salamanca 1997; McNamara 1997).  
Furthermore, the bright, blue continuum in two objects (the A2597
and A1795 CDGs)
lies along their radio sources in a manner similar
to the high redshift, FR II radio galaxies, but on a smaller scale.  
Although these two objects
share some of the alignment properties seen in FR IIs, 
they are noteworthy in several 
respects.  They reside at the centers of bright,
cluster-scale, thermal X-ray emission that has been
interpreted as the signature of a large reservoir of cooling gas
(i.e. a ``cooling flow'').   They harbor less-luminous, 
FR I radio sources whose
1.4 GHz radio powers are less than $10^{26}~{\rm W~Hz}^{-1}$, and
their aligned blue continuum, or ``blue lobes'', is preferentially
associated with their radio lobes, rather
than their radio jets. Finally,
A2597 and A1795 are relatively nearby at redshifts of 0.082 and 0.064
respectively.  A redshift or radio luminosity dependence on the degree
and frequency of strong alignments would suggest an 
active evolutionary phase of 
radio galaxies that occurred at a special cosmic epoch or under
special circumstances.  

In order to explore the emission mechanism of the radio-aligned 
continua  in the A2597 and A1795 CDGs, we have obtained U-band polarimetry
of the blue continuum along their radio sources.
Polarimetry can be used to discriminate between 
the two favored emission mechanisms: highly polarized scattered light from a hidden
active nucleus 
(Sarazin \& Wise 1993; Crawford \& Fabian 1994; Murphy \& 
Chernoff 1993; Sarazin \etal\ 1995) and the unpolarized light from
star formation (MO93; De Young 1995).  The scattering model was proposed 
in part because of its success in explaining
the often highly polarized continuum in the distant
FR II radio galaxies (di Serego Alighieri \etal\
1989; Scarrott, Rolf \& Tadhunter 1990; Jannuzi \& Elston 1991;
Tadhunter et al. 1992; di Serego Alighieri, Cimatti \& Fosbury 1993;
Antonucci 1993; Jannuzi 1994; Jannuzi
\etal\ 1995; Dey et al. 1996;  Cimatti \etal\ 1997), and in
response to the paradigm that seeks to unify FR I radio
sources and BL Lac objects (Urry and Padovani 1995).
The scattering models are appealing in the cases of A2597 and A1795
because the dense and dusty environments associated with their
cooling flows provides a suitable scattering medium, were the FR I radio
sources indeed BL Lac objects seen at an oblique angle to the line of sight
(Padovani \& Urry 1990; Urry and Padovani 1995; Sarazin \&
Wise 1993; Murphy \& Chernoff 1993).  Were the blue lobes shown
to be highly polarized with the electric vectors orthogonal to
the radio axis, the result could be interpreted as a significant
step toward verifying such a ``unified scheme.''

We have addressed these issues using sensitive U-band 
polarimetric observations of the blue lobes. 
Although these galaxies are intrinsically faint at U,
and the throughput of the telescope and
polaroid filter at U are low, the U-band offers the largest contrast
between the blue continuum and the red background population of
the cD galaxy.
The U-band is therefore most sensitive to the blue continuum
and least prone to error when estimating the
amount of presumably unpolarized background light that would
dilute any polarized signal from the blue lobes.
The background starlight limits the accuracy of the 
polarization measurements (\S4.1).
Therefore, minimizing the amount of dilution by starlight 
is critical to maximizing the sensitivity to low
levels of polarized light.  In addition, shorter wavelength observations are 
more useful for comparison to the high redshift FR~II radio galaxies,
where R and I-band observations correspond to rest wavelengths in the U-band.

In a study similar to that presented here for A2597, we 
found A1795's aligned continuum, or blue lobes, 
to be unpolarized (McNamara \etal\ 1996a).
In subsequent papers the lobes were shown to 
be resolved into what appear to be star clusters using
images obtained with the Hubble Space Telescope (McNamara \etal\ 1996b;
Pinkney \etal\ 1996), clearly demonstrating the emission
mechanism for the lobes in A1795 to be a population of young stars.
In this paper we present the results 
of a similar polarimetric study of the blue lobes in the
A2597 CDG.  
%We find its lobes to be unpolarized and argue,
%similarly to the case of A2597, that its blue lobes are
%regions of star formation that that were probably
%triggered by the radio source.

%All distance- and time-dependent quantities
%discussed in this paper assume $\ho=50 ~\hounit$, and $q_0=0$.

\section{Observations}

The images were obtained with the Mayall 4 meter telescope of the Kitt Peak
National Observatory 
during the nights of 8, 9, and 10 November, 1996.  We used the
Tektronix $2048\times 2048$ pixel (scale$=0.47$ {\rm arcsec}/pixel) CCD detector mounted
at the prime focus. The Q and U Stokes parameter
images were
constructed from a series of exposures obtained through a combination 
of a U-band filter attached to a copper sulfate
blocking filter, and one of four Polaroid
filters (HNP'B sheet Polaroid) with transmission axes at 
$0\deg ,~ 45\deg ,~ 90\deg ,~{\rm and } ~135\deg$.
%The HNP'B sheet polaroid plus U-band filter provided a throughput 
%of $XX\%$. 
We obtained 9 or 10 CCD images exposed for 800 seconds at each
position angle for a total of 31,200 seconds of integration time.  
The data were taken during transparent but 
unphotometric conditions.  Further details of our observing and data reduction
technique are presented in McNamara \etal\ 1996.  

\section{Properties of the Central Dominant Galaxy}

In Figure 1 we present a composite image of the A2597 CDGs blue lobes
embedded in the smoothed, grayscale contours of the U-band image of
the galaxy.  The superposed white contours show the 8.44 GHz radio
emission mapped with the VLA, presented earlier
in Sarazin \etal\ (1995).  The composite
U-band image was constructed first by subtracting the model for the background
galaxy discussed in Section 4.1 from the summed,
31.2 ksec U-band image, leaving the
bright blue lobes in residual. The residual image was then multiplied
by an arbitrary factor,
and the U-band image was smoothed with a 4 pixel FWHM Gaussian kernel.
The residual and U-band images 
were scaled logarithmically, added,
and displayed in grayscale. This rendering allows the brightest
and bluest regions in black to be seen against the background
galaxy contours in gray and the radio source in white.  The structure seen in
earlier U and I data by MO93 and Sarazin \etal\ (1995) is clearly
seen in our new U-band data, although the pixels in the new data
subtend a larger angular size, and details on scales smaller than
$\simeq 2$ arcsec may be unreliable.  

We will not repeat the detailed 
discussions of the galaxy's properties  presented in MO93 and 
Sarazin \etal\ (1995), but we will mention a few salient
properties pertaining to this discussion.  The radio jets are seen in Figure 1 to emerge from
the nucleus in a north-east/south-west direction along the minor
photometric axis of the galaxy.  The bright blue lobes, shown in black,
are located near the radio lobes. They are brightest and bluest 2--3 arcsec
from the nucleus, where the colors of the blue lobes are 0.7--1.0 
magnitudes bluer in $U-I$ than the colors of a normal
giant elliptical at that radius.
The most striking indication that the radio source and
matter associated with the blue lobes are interacting is
the sharp bend in the radio structure to the south-west, where the
radio lobe seems to be expanding and bending at the location
of the southern blue lobe.  O'Dea, Baum, and Gallimore (1994)
discovered H I in absorption against the radio lobes with 
a broad, $\sim 410~\kms$ FWHM, turbulent velocity structure
whose mean velocity is consistent with the mean velocity
of the galaxy.  Sarazin \etal\ (1995) suggested that the radio 
jets may have been deflected
and the radio lobes disrupted as the outwardly moving radio plasma
collided with the H I clouds. The H I clouds may be associated
with the bright emission-line nebula embedded in the inner 20 kpc
of the galaxy (O'Dea, Baum, and Gallimore 1994).

\section{Polarization Analysis }

We determined the degree of polarization of the U-band light emitted from the
entire central blue region and from the blue lobes individually.
To do so, we extracted the net fluxes in these regions 
 using
circular synthetic apertures applied to the sum of the CCD frames
for each transmission angle, after subtracting the sky background from
each of the CCD frames.
The Stokes flux for each aperture was computed by taking 
flux differences
between transmission axes, $S_\theta$, as $Q=S_{0\deg}-S_{90\deg}$,
$U=S_{45\deg}-S_{135\deg}$.  The normalized Stokes 
flux was found as $q_{\rm
n}=Q/(S_{0\deg}+S_{90\deg})$ and $u_{\rm n}=U/(S_{45\deg}+S_{135\deg}$),
where the degree of instrumental plus total polarization was found as 
$P=\sqrt{q_{\rm n}^2 + u_{\rm n}^2}$.  The degree of total polarization  (background stars
plus blue lobes) 
of the flux at each aperture position on the galaxy was found by
subtracting the mean instrumental polarization determined using the presumably
intrinsically unpolarized flux for eight reference objects in the field surrounding the
central galaxy as $P_{\rm tot}= P_{\rm gal}-P_{\rm ref}$.

A summary of the polarization measurements and the sizes and locations of the
apertures are given in Table 1.  Columns
1--3 give the locations of the aperture centers with respect to the
nucleus, defined as the peak in the U-band flux.  The offsets from the
nucleus are given in column 2, and the position angles measured from
north through east are given in column 3.  Column 4 lists
the diameters of the circular apertures.  
Column 5 lists the total polarization found in each
aperture, and column 6 gives the RMS deviations about the 
mean instrumental polarizations for the eight reference objects. Column 
7 lists the upper limits to the degree of polarization of the blue
lobe light in each aperture,
after correcting for  background light.  The procedure we used to 
derive the data listed in column 7 is described in Section 4.1.
In column 8, we list the net polarization after accounting
for dilution by the stellar background (i.e. column 7) and 
nebular emission (\S4.2).

An inspection of columns 5 and 6 in Table 1 shows no significant 
total polarization for the large central aperture or the 
smaller apertures circumscribing the blue lobes.   
The upper limits to the degree of polarization of the 
total light from the lobes (lobes $+$ galaxy) is less than $2\%$, based on
the scatter in the measured degree of instrumental polarization of the reference 
objects.  The statistical error in the each of the
Stokes parameters, which included variations in the U-band sky
background ($\leq 1\%$) and photon statistics ($\ll 1\%$), was found to be
less than $1\%$.  
%A nearly identical polarization of $2\%$ was found for a number 
%of presumably unpolarized galaxies in the field.  This offset reflects the
%systematic error associated with registering and scaling
%the images and perhaps some intrinsic polarization across
%the field of view imparted by aligned dust grains along
%the line of sight through our Galaxy.  

\subsection{Stellar Background Model}

In order to determine the upper limits to the intrinsic polarization 
of the blue lobes, we modeled and removed the contribution of 
presumably unpolarized starlight from the galaxy.  
The background galaxy model was constructed by first measuring the
U-band radial surface brightness 
profile of the galaxy.  This profile was constructed 
by extracting fluxes from elliptical annuluses
with shapes defined by the I-band major axis position angles and 
isophotal ellipticities, based on data from McNamara \& O'Connell (1993). 
The radial surface brightness 
model was constructed by fitting a straight line to the U-band radial surface brightness 
profile in magnitudes per square arcsec against semimajor axis to the $1/4$-power.  
This $R^{1/4}$-law profile was fit to the data at
radii between 13 and 20 arcseconds, well beyond the
the blue central region of the galaxy where the colors are apparently
typical of a normal cD galaxy.  The surface brightness
profile is shown as solid dots, and the fitted $R^{1/4}$-law profile
is shown as a solid line in Figure 2.  The model profile is
extrapolated inward in order to construct the model of the older
background population at the location of the lobes.  The departure of
the observed surface brightness profile above the $R^{1/4}$-law profile
is clearly seen in the inner 8 arcsec or so where the excess blue
light and line emission are observed (McNamara \& O'Connell 1993; Sarazin \etal\ 1995; 
Cardiel, Gorgas, \& Aragon-Salamanca 1997).  The $R^{1/4}$-law 
profile overestimates the contribution of background light in
the inner arcsec or so of the galaxy. This should have little
effect on the estimated contribution of background light at
the locations of the blue lobes.  The model surface brightness 
profile was then applied to the corresponding semimajor axis
locations in the artificial image of the galaxy whose shape was
identical to the mean isophotal shape of the I-band image of
the galaxy.

The contribution of background light at the locations of the apertures 
placed on the image of the real galaxy was found by measuring the flux in the model
galaxy using identical apertures and locations as for the real galaxy
and reference objects (i.e. Table 1).
The Stokes parameters and degree of polarization of the net flux
from the lobes after subtracting the galactic stellar model background,
$f_{\rm M(r)}$, were found as $q=Q/(S_{0\deg}+S_{90\deg}-{f_{\rm M(r)}})$,
$u=U/(S_{45\deg}+S_{135\deg}-{f_{\rm M(r)}})$, $P=\sqrt{q^2 + u^2}$,
and $P_{\rm *}=P_{\rm gal}-P_{\rm ref}.$

The values of $P_*$ found for each aperture are given 
in column 7 of Table 1.  They show that the three sigma upper limit to
the degree of polarization of the net flux from the lobes is
less than $5\%$. The three sigma limit was computed
by adding an offset to the surface brightness model profile 
equal to three times the statistical error in
the zero point of the $R^{1/4}$-law fit to the data, and then
following through with the analysis of the
model data as described above.  The RMS error associated with the measurement
of the total polarization prior to considering the background model is
less than $2\%$.  Therefore, dilution by ambient starlight
surrounding the blue lobes is the factor limiting the precision
of the polarization measurement.  The $3 \sigma$ upper limit to 
the total polarized flux (lobe + galaxy) of both lobes at $U$
is $<1.7 \times 10^{-15}~{\rm erg~cm}^{-2}~{\rm s}^{-1}$. 
This figure was estimated using
calibrated photometry (McNamara \& O'Connell 1993; Sarazin \etal\ 1995),
and the upper limit of 2\% to the degree of polarization of the
total light at $U$ presented here. 

\subsection{Nebular Emission}

Diffuse Nebular line emission in the vicinity of the A2597 CDGs 
blue lobes is
quite strong (Voit \& Donahue 1997), and the contributions of
diffuse nebular continuum and line radiation to the U-band could be
significant.  Unpolarized nebular radiation would dilute a 
polarized signal.

In order to determine the fraction of the U-band color  
excess that may be attributable to unpolarized nebular continuum,
we estimated the amount of recombination radiation from
hydrogen and helium, two photon emission,
and bremsstrahlung radiation that would be expected for the conditions
in A2597.  We calculated these emissions 
relative to the strength of the H$\beta$ emission line using
the Case B emission coefficients tabulated in Aller (1984) and
Osterbrock (1974).  
The contribution of nebular continuum to the $U-B$ color excess 
in the vicinity of the blue lobes can be described then as, 
\begin{equation}
\Delta (U-B)\simeq -2.5~{\rm log}\left[1 + {\epsilon(T,\Delta\lambda_U) EW({\rm H}\beta) \over \Delta \lambda_U }\right].
\end{equation}

In this expression, $EW({\rm H}\beta$) is the equivalent width
of the H$\beta$ emission feature, $\Delta \lambda_U\simeq 600$ \AA\ is the approximate
effective width of the $U$ passband, and 
$\epsilon(T,\Delta\lambda_U)$ is the ratio of 
the strength of the nebular continuum averaged over the $U$ passband 
to the H$\beta$ line flux.  We assumed half solar abundances, a gas density of
$200 ~{\rm cm}^{-3}$, and an ion temperature of 10,000 K,
appropriate to the nebula surrounding
the blue lobes (Voit \& Donahue 1997).
The strength of the nebular continuum does not depend significantly
on the ion density for the densities of interest here, but
it does depend somewhat on ion temperature.  
We have
adopted $\epsilon=2.8$ to be consistent with the Voit \& Donahue
(1997) analysis.  We are unaware of a tabulated measurement in the literature
of the H$\beta$ equivalent width for the A2597 CDG in the vicinity
of the blue lobes.  Therefore, we estimated the H$\beta$ equivalent
width to be 
$EW({\rm H}\beta)=28$ \AA\ in the nucleus using the tabulated H$\beta$
flux and the continuum spectrum given
in Voit \& Donahue (1997).  Cardiel \etal\ (1998) measured
the radial profile of the H$\beta$ equivalent width and similarly found it
to be 28 \AA\ in the nucleus, after correcting for H$\beta$
absorption.  They found that equivalent width decreases with radius to
$\simeq 15-23$ \AA\ at the radius of the blue lobes.  Unfortunately
Cardiel's measurements were made using a slit placed perpendicular
to the blue lobes along the major axis of the galaxy, so we do not
know the precise value of the  H$\beta$ equivalent width at the
location of the lobes. 
We find that the color excess at the location of the blue lobes that can be attributed to
unpolarized nebular continuum is $\Delta (U-B)\gae -0.07 ~{\rm to}~ -0.11$
magnitudes.  This color excess corresponds to H$\beta$ equivalent widths of 20\AA\ and
30\AA\ respectively, which bracket the values in the nucleus and at
the radius of the blue lobes.  The upper bound on
the color excess contributed by unpolarized nebular emission
should be reasonable because the
nebular emission is centrally concentrated (Heckman \etal\
1989), and our estimate exceeds the nuclear value.
Nonetheless, our estimate
should be taken with due caution, absent a direct measurement, as the H$\beta$ equivalent width
could be larger near the blue lobes than we have assumed.
The observed total color excess in A2597's lobes is $\Delta (U-B)\simeq -0.6$ magnitudes
(McNamara 1997; Cardiel, Gorgas, \& Aragon-Salamanca 1997).
Therefore, we expect the unpolarized nebular continuum to comprise
only $\simeq 10-15$\% of the observed total color excess.

The uncertainty in the measurement of the total color excess is
caused primarily by the unknown dust distribution within the nebula. 
Voit \& Donahue (1997) found $\simeq 0.3$ magnitudes of
extinction at $U$ based on departures of observed Balmer line
ratios from Case B predictions.  This extinction should not affect appreciably
our estimate of the percentage of the color excess that is
contributed by nebular and stellar continuum unless the gas, dust,
and young stars are distributed differently.  The existing data 
do not allow us to determine whether this is so.

The equivalent width of A2597's  [O II]$\lambda$3727 emission feature 
is $\simeq 247$ \AA\ in the nucleus (Voit \& Donahue 1997).
However, assuming the [O II] equivalent width decreases
with radius proportionally to the H$\beta$
feature, we expect the [O II] equivalent width at the location
of the lobes to be roughly between 130--200 \AA .
The [O II] feature's location is redshifted to $\lambda$4034 \AA\ in the
laboratory frame, where the throughput of the U-band filter is about
5\%. We estimate the contribution of [O II] emission to the
$U$ band at the location of the lobes to be between 0.01--0.02 magnitudes, or about 3\% of the
color excess.  

In summary, we estimate the total nebular contribution to the $U$-band
color excess, including [O II] emission plus nebular continuum,
to be between 13--18\%.  This continuum would dilute a polarized signal
from the blue lobes and increase the
upper limits to the degree of polarization by one percent or
less (i.e. $P_{\rm *,neb}$, Table 1, column 8).

\section{Radiation Mechanisms for the Blue Lobes}

\subsection{Interpretation of the Polarization Upper Limits}
\label{sec:radiation_scatter}

We now calculate the expected polarization of the blue lobes in
A2597 if they are due to electron scattering or dust scattering,
and compare to the observed upper limit.
The expected linear polarization of the blue lobes in the A1795 CDG
due to electron scattering by the cooling flow was calculated by
Sarazin \& Wise (1993) and was
discussed in detail in McNamara et al.\ (1996a).
The blue lobes in A2597 are similar in many respects, and our analysis
follows that for A1795 (McNamara et al.\ 1996a).
We assume that the blue lobes are due to scattering of beamed radiation.
We consider beamed rather than isotropic emission
because no strong nuclear point source is seen in A2597
(Crawford \& Fabian 1993; Sarazin \& Wise 1993). 
The model polarization was calculated for the scattered light only,
without dilution by the background galaxy light.

First, we consider the possibility that the blue lobes are due to electron
scattering.
We assume single electron scattering, as the observed electron scattering
optical depth of the cooling flow is small (Sarazin et al.\ 1995).
In this limit, the polarization is independent of the scale of the
electron density of the cooling flow.  Of course, the polarization is
always independent of the flux of the anisotropic nuclear source."
The assumptions included in our scattering model are identical
to those discussed in McNamara et al. (1996a), with the
exception of the opening angle of the scattering cones.
We assume that the electron density, $n_e$, in the cooling flow varies
with radius, $r$, as $n_e \propto r^{-1}$, which gives a reasonable
fit to the observed X-ray surface brightness at small radii
(Sarazin et al.\ 1995).

For this calculation, we assume that anisotropic radiation from
the central nucleus is conical, initially unpolarized, and uniformly
illuminated.
The polarization is calculated including the effects of averaging along
the line-of-sight through the beam and averaging the azimuthal
polarization across the projected width of the beams.
The polarization depends slightly on the observed width of the lobes.
We estimate an angular half-width of
$\phi_{max} \approx 35 - 45^\circ$ for the NE lobe and 
$\phi_{max} \approx 25 - 35^\circ$ for the SW lobe.
We adopt $\phi_{max} = 35^\circ$, but note that
the resulting polarization is not strongly dependent on this assumption
or any of the other assumptions, as shown in Sarazin \& Wise (1993)
and McNamara et al.\ (1996a).
The predicted polarization does depend
strongly on the angle $\theta$ between our line-of-sight and
the central direction of the beams.
(See Figure 1 in Sarazin \& Wise [1993] for the definitions of 
the angles.)
For each value of the angle $\theta$, we determine the
angular width of the beams $\theta_b$ which is consistent with the
observed width $\phi_{max}$ of the blue lobes in A2597.
The existence of distinct lobes and the fact that the nucleus is
not extremely bright both require that our line of sight be
outside of the beams, so that $\theta_b < \theta$.

Figure 3 shows the predicted polarization (solid line)
of the electron--scattered lobe light, $P$, as a function of the angle of
the beams to the line-of-sight, $\theta$.
The observed upper limit of $P < 6$\% is shown as a long dashed horizontal
line.
The observed upper limit is only consistent with the prediction
of the simple electron scattering model if $\theta < 20^\circ$.
The probability, $P_o$, that the beam would be randomly oriented this close to our
line-of-sight is $P_o< 6$\%.
(Note that for small polarizations, the probability and the polarization are
always nearly equal.) 
Thus, the consistency of the observed upper limit on the polarization
of the lobes with the simple electron scattering model would require
an unlikely near alignment of the beams with our line of sight.
A very similar result was found for the blue lobes in A1795, where the
limit on the orientation angle and probability were $\theta < 22^\circ$
and $P_o < 7$\%.
The strong upper limits on the polarization of the lobes makes it unlikely
that they are due to electron scattering.
While a single case might be the result of an unlucky alignment, it seems
very unlikely that both could be explained this way.

This calculation of the polarization in Figure 3
assumed that the emission from the nucleus of the AGN was unpolarized.
If the nucleus contains a BL Lac object, then the nuclear emission
might itself be highly polarized.
In most situations, this will increase the polarization of the scattered light
(Sarazin \& Wise 1993).
However, for certain very restrictive choices of the angles, the polarization
can decrease, although this requires that the parameters be at least as
finely tuned as in the case for scattering of unpolarized radiation
(McNamara et al.\ 1996a).

We have also calculated the predicted polarization for scattering by dust.
We assumed the dust scattering properties given by White (1979) for
the standard MRN model for interstellar dust.
The predicted polarization for dust scattering is shown by the short
dashed curve in Figure 3.
Unlike electron scattering, dust scattering is not symmetric between
forward and backward scattering, and thus the two lobes would have
different polarization.
Since neither of the two observed lobes shows any evidence of polarization,
the curve shown is the maximum of the polarizations of the two lobes.
At small and large values of $\theta$, the polarization is larger for
the back lobe (the one further away from us, for which the scattering
is predominantly back scattering).
At intermediate angles ($37^\circ \le \theta \le 61^\circ$) the
polarization of the front lobe is higher.
At small angles ($\theta < 37^\circ$), the predicted polarization due to
back scattering is radial for the more strongly polarization lobe.
At all other angles, the predicted polarization is azimuthal.
(The polarization due to electron scattering is always azimuthal for
an initially unpolarized source.)

For both A1795 and A2597, the lack of observed polarization in the lobes
and the fact that they are not very strongly asymmetric in their brightness
makes it unlikely that they are due to dust scattering.
The polarization produced by dust scattering is smaller than that produced
by electrons.
As a result, the limit on the angle $\theta$ is weaker.
The observed limit on the polarization of less than 6\% implies that
$\theta \le 22^\circ$ or $30^\circ \le \theta \le 44^\circ$,
for which the probability is $P_o \le 23$\%.
However, dust scattering is not symmetric forward to back, and the same
conditions which would lower the polarization (small $\theta$)
would produce rather
asymmetric lobes, with a ratio of fluxes which would exceed $\ge$6.
% ?? Brian:  Do you still agree with this number?
While the two lobes are certainly not symmetric, the ratio of their
surface brightnesses is $\la$4
(Sarazin et al.\ 1995).
We have also determined the predicted polarization due to dust
scattering in A1795
(McNamara et al.\ 1996a).
There, the 3-$\sigma$ upper limit polarization leads to an upper limit on
the angle of $\theta \le 46^\circ$, a probability of
$P_o \le 31$\%, and a flux ratio of $\ge$5.8.

Given these limits on dust scattering and the stronger limits on
electron scattering, it is improbable
that the lobes are due scattering of beamed light from the nuclei of the
galaxies.  The absence of a polarized signal or of a detailed correspondence 
between the radio and optical morphologies renders synchrotron radiation 
an unlikely emission mechanism.  
Furthermore, Compton scattering of microwave background photons by relativistic electrons 
associated with the radio source (c.f. Daly 1992) is incompatible 
with the object's proximity and radio power, and optical bremsstrahlung radiation from the diffuse X-ray source would
be too week to explain the blue lobes for the observed gas densities
in A2597 (Sarazin \etal\ 1995).

Finally, our polarization measurements probe directly the 
paradigm that seeks to unify FR I radio
sources and BL Lac objects (e.g. Urry and Padovani 1995).  
The A2597 and A1795 CDGs reside in hot cluster atmospheres
with central gas densities of $\sim 10^{-1}~{\rm cm}^{-3}$.
Were the CDGs to contain typical BL Lac nuclei,
$\sim 1\%$ of the anisotropically
emitted radiation from the BL Lac would be scattered off of electrons
into the line of sight, which should be detectable (Sarazin \& Wise 1993).
Furthermore, the alignment of the blue lobes with the 
radio sources (McNamara \& O'Connell 1993) is consistent with the
scattering hypothesis, which prompted Sarazin and Wise to 
investigate its feasibility. Their scattering model assumes
anisotropic nuclear emission directed obliquely to the line of sight,
with luminosities comparable to
a typical BL Lac object ($L=10^{47} \ergsec$) in the spectral range
$\Delta \nu=10^8-10^{18}$ Hz.  The scattering medium was assumed
to be an electron gas of comparable density to the X-ray-determined
values at the centers of the cooling flows.  The predicted $U$-band surface 
brightness of the scattered light matched closely the observed surface
brightness of the lobes in A2597 and A1795 found by 
McNamara \& O'Connell (1993).  However, the remaing critical question
was the degree of polarization of the $U$-band lobe emission,
which should be $> 8\%$ were the blue lobes scattered light.  The
polarization measurements presented here were intended to test
the scattering model.   Our upper limits are then inconsistent with the
scattering model and do not support the FR I--BL Lac unification 
paradigm to the extent that the assumptions made
by Sarazin \& Wise (1993) are reasonable.

\subsection{Radio Triggered Star Formation}

Following on the previous section, we conclude
that the radiation from blue lobes is most likely primarily continuum from young, blue stars.
This interpretation receives further support by a recent analysis of the
nebular emission surrounding the lobes (Voit \& Donahue 1997).
The spatial correlation between the blue lobes and radio lobes shown
in Figure 1, McNamara \& O'Connell (1993), and Sarazin \etal\ (1995) 
suggests that the
blue lobes and the radio source are related either by chance:
the radially expanding radio jets happened upon the dense
clouds associated with the blue, star-forming regions, or causality:
the star formation was triggered by the radio source.
A mere coincidence seems unlikely.  There is one other obvious
example of this phenomenon in the A1795 CDG, which, as 
A2597, was identified
among the roughly two dozen or so clusters with large cooling flows
whose CDGs have been well imaged from the ground (McNamara 1997).
The bluest and presumably the youngest regions of star formation
have been shown, using HST imagery, to lie along the edges of the radio lobes
in A1795 (McNamara \etal\ 1996; Pinkney \etal\ 1996).  Although the 
bluest regions in A2597 are located near the
radio lobes (i.e. Figure 1), HST images show that
they do not correlate strongly with the edges of the radio lobes
(Koekemoer \etal\ in preparation), as is seen in A1795.  Although this would
render A2597 a less compelling case for radio-triggered star
formation, the mechanism by which star formation is induced 
is poorly understood, and predictions based on such models are
uncertain.

De Young (1995) presented a model to explain the lobes 
in A1795 as a burst of star formation
triggered by the rapid collapse of cold clouds
compressed by shocks along the expanding radio jets.
This general scenario is consistent with
bends in the radio sources of both objects occurring
near regions of dust extinction, H$\alpha$ emitting gas, and
in A2597, near H I absorption clouds (McNamara \etal\ 1996; O'Dea
\etal\ 1994; Koekemoer \etal\ in preparation), which indicates
that the radio sources are interacting with cold gas clouds.
However, De Young's model does not readily explain the 
location of the bluest, and presumably the youngest
star clusters  along the edges of the
radio lobes in A1795, rather than along the edges of the jets.  

The weaker correlation between the bluest regions and the radio
source in A2597 is not necessarily surprising.  
A strong correlation between the edges of the radio source
and the sites of star formation should be short lived
(McNamara \& O'Connell 1993; De Young 1995; McNamara \etal\
1996a; Cardiel, Gorgas, \& Aragon-Salamanca 1997).
The nebular and H I gas velocities of a few hundred
kilometers per second near the
lobes are disordered (Heckman \etal\ 1989; O'Dea \etal\ 1994).
Assuming the star formation is fueled by cold gas with
a similar velocity structure,
the stellar lobes should disperse quickly, and certainly in less than 
the local free-fall time of a few tens of Myr.  
Therefore, it is possible that we have caught the burst
of star formation in A2597 several Myr after it was
initiated by the radio source, and stars have begun to 
disperse as the radio source expands outward.   

The ground-based photometry and radio data are not capable of
pinpointing the relative ages of the radio source and 
regions of star formation to reliably test this hypothesis.
Nonetheless, the short dispersal timescale would be
consistent with both the lower limit on the radio age of
0.5 Myr (Sarazin \etal\ 1995) and with the colors for a burst of star
formation that occurred roughly 5 Myr ago with the local initial mass function and solar
abundances (McNamara \& O'Connell 1993).  The stellar mass of the
blue lobes composed of such a population would be $\sim 10^8~\msun$, 
which implies a total star formation
rate in both lobes of roughly $20 ~\msunyr$. This star formation rate
and stellar mass does not include adjustments upward for extinction,
and adjustments downward for nebular emission (\S 5.3).  
Including a $\Delta(U-B)\simeq -0.3$
magnitude extinction correction, and a $\Delta(U-B)\simeq +0.1$
magnitude nebular emission correction, the young stellar mass and
star formation rate would increase by about 68\%.  This systematic
increase does not exceed the uncertainties of the estimate of the 
uncorrected mass, 
which matches reasonably well with the mass of neutral hydrogen of $\sim
7\times 10^7\msun $ estimated from
the VLA H I absorption measurements (O'Dea \etal\ 1994).

\section{A Comparison Between the Polarized Luminosities and Radio
Power for Radio Aligned CDGs and High Redshift Radio Galaxies}

In this section we explore the similarities and differences between the
alignment properties of the high redshift radio
galaxies (HzRGs) 
and those in A2597 and A1795.  We wish to determine whether
they are fundamentally different types of object,
or whether they are similar albeit on much different spatial
and energy scales.  In order to do so, 
we will begin by contrasting a few of their relevant properties, most
importantly their polarized luminosities and radio powers.

The blue optical continuum found along the radio sources
in HzRGs is often polarized at levels of several to greater than ten percent.
The electric vectors are, in general, nearly perpendicular
to the radio and optical continuum axes.  In addition, they have
strong, extended, nebular line emission near their radio sources.
Based on these facts, a consensus
has emerged that would explain the aligned optical continuum
as being primarily scattered light from a powerful, 
misdirected active nucleus or QSO, plus a smaller but
significant contribution of nebular continuum
(e.g. Dey 1998; Cimatti \etal\ 1997; Dey \etal\ 1996; Stockton,
Ridgway, \& Kellogg 1996; Dickson \etal\ 1995).  While this may be
the case in the majority of aligned HzRGs with reliable
polarimetry, it is not always true.
For example,  the aligned 
components in 3C 285 (van Breugel \& Dey 1993) 
and 4C 41.17 (Dey \etal\ 1997) are unpolarized. Their aligned
continua appear to originate from young stellar populations.
Furthermore, even in those objects with a high degree of
polarization, star formation at some level cannot always be excluded
(Cimatti \etal\ 1996).
Therefore, there appear to be at least three 
physical processes corresponding to three 
emission mechanisms that contribute to the alignment effect in  HzRGs:
scattered light, nebular continuum, and star formation.

In contrast, the alignment seen in A2597 is associated with a smaller, 
lower power, FR I radio source.
Its radio-aligned, $U$-band optical continuum 
is composed primarily of light from a young, $\sim 10$ Myr old
stellar population ($\gae 80\%$), a 
small contribution of nebular emission ($\sim 10-20\%$),
and at most a minor contribution of scattered light.
The situation is similar in the A1795 CDG (McNamara \etal\ 1996a). 
In addition, the blue optical emission found near the radio lobe
of the FR I radio source PKS 0123-016A, the well-known ``Minkowski's Object,'' 
is primarily from a young stellar population (van Breugel \etal\ 1985). 
Therefore, both star formation, and to a small degree nebular
emission, contribute to the aligned
optical components in the FR Is.  However, we are unaware of 
any evidence of scattered radiation playing a major
role in producing the aligned optical components of an FR I radio galaxy.

%It is not clear that the differences in composition
%of the aligned components in FR I and FR II radio galaxies
%imply that they are fundamentally different objects.

The environments of the FR I radio sources in the
A2597 and A1795 central cluster galaxies
and HzRGs that exhibit the alignment 
effect are similar in at least three general but important
respects.  Both types of radio source seem to be found in elliptical
host galaxies (Rigler \etal\ 1992; Cimatti \etal\ 1994; Dey 1998).   
Second, one can infer by the presence of radio sources  
that they contain a central engine that powers their radio sources.
Third, by virtue of the presence of strong nebular line emission
and recent star formation, the host galaxies must harbor 
reservoirs of cool gas (e.g. McNamara 1997 and references therein).
A2597 and A1795 are dissimilar to the powerful HzRGs, 
lacking evidence for bright, blue, unresolved continua in their
nuclei (e.g. McNamara \etal\ 1996b; Pinkney \etal\ 1996) or 
broad emission lines (Heckman \etal\ 1989).

The characteristic that perhaps best distinguishes between
A2597, A1795, and the HzRGs is radio power.
We calculated the radio powers for several 
high redshift 3C radio galaxies observed with the Keck Observatory 
(e.g. Dey 1998).  The radio powers
in the restframe 1.4 GHz bandpass, $P_{1.4}$, were found by interpolating 
between measured flux densities bracketing the restframe
frequency $\nu=1.4/(1+z)$ GHz. We have assumed $\ho=50~\hounit$ and
$q_0=0.5$ throughout our calculations.  The computed flux densities
and radio powers are presented in Table 2.  In columns 1 and 2
we list the radio galaxy name and redshift.
Column 3 lists the restframe radio flux
density; column 4 lists the restframe radio
power; column 5 lists the degree of polarization in the
restframe U passband, determined from spectropolarimetry
obtained with the Keck Observatory.  In column
6 we list the restframe U-band polarized continuum luminosity.  The polarized
luminosity was determined as 
\begin{equation}
L(U)_{\rm pol}=4\pi D_{\rm lum}^2f_{\lambda 3600}\Delta \lambda_U P(U),
\end{equation}
\noindent
where $f_{\lambda 3600}$ is total flux at 3600 \AA , and
$P(U)$ is the degree of polarization at 3600 \AA\ in the
rest frame.  The effective rest frame $U$
passband is assumed to be $\Delta \lambda_U = 600$ \AA, and
$D_{\rm lum}$ is the luminosity distance to the radio galaxy.
In column 7 we index the references to the data.
We have included entries in Table 2 for A2597 and A1795 for
comparison.  Because we
computed the polarized luminosities and radio powers of the HzRGs 
in the restframe U-band, they can be compared directly
to those for A2597 and A1795.  

The radio power is plotted against the restframe U-band polarized continuum 
luminosity in Figure 4.  
Note the difference of a factor of several thousand 
in radio luminosity between the A2597, A1795, and the aligned
HzRGs.  In addition, the polarized luminosities of the HzRGs 
are $\sim 600\times$ larger than the upper limits for the FR Is.
In all likelihood, the much larger radio powers in HzRGs are
the result of a much stronger AGN.  Consequently, the HzRGs are
capable of producing a much higher scattered light 
intensity for a given ambient electron or dust density.
Now we ask whether
we would have detected any polarized flux in A2597 and A1795
if the polarized luminosity scales in proportion
to the radio power.  The solid line in Fig. 4 represents 
$L(U)_{\rm pol}\propto P_{\rm rad}$.  The line is scaled to
the median of the 3C radio galaxies. We can see that
the ratio of polarized luminosity to radio power for the
HzRGs extrapolated downward to the radio power of the aligned 
FR Is falls a factor of about 30 below the upper limits for
A1795 and A2597.  
Therefore, if $L(U)_{\rm pol}\propto P_{\rm rad}$ our polarimetry
for A2597 and A1795 would not have detected the polarized signal. 
The composition (electrons, cold gas, dust), 
density, and spatial distribution of the interstellar medium of the host 
galaxy, and the orientation of the AGN are factors 
that would contribute scatter to the relationship plotted in Figure 4.  
However, by using several, well-observed 3C galaxies, 
we have attempted to average over these
factors.  In any case, when the line is scaled
to the upper envelope of the group of 3C galaxies
in Figure 4, the A2597 and A1795 upper limits remain well above 
the line.  Therefore, we conclude with reasonable confidence
that were the AGN in the A1795 and A2597 CDGs the dwarf siblings of the
aligned HzRGs, we would not have detected their polarized 
signals.  This
result depends on the adopted cosmology in detail, but the 
conclusion is unaffected.

Our conclusion rests uncomfortably on an extrapolation of 3--4 orders
of magnitude in radio power and a factor of 600 in polarized
luminosity.  It would be reassuring to find and explore 
alignment effect radio galaxies that lie between the FR I and
HzRGs in Figure 4.
Furthermore, if A1795 and A2597 indeed contain weak AGN similar
in nature to the HzRGs, a polarized signal of 
$\sim 2\times 10^{-16}~{\rm erg~cm^{-2}~sec^{-1}}$ should  
be present, if our extrapolation to low radio power is
correct.  This flux is roughly an order of magnitude
below our limits for A1795 and A2597.  However, a polarized flux at this 
level may be detectable near the nucleus using high resolution 
space polarimetry, or ground-based polarimetry using a
large telescope in excellent seeing.

Finally, we highlight earlier remarks that 
pure Thompson scattering models imply the presence
of large cooling flows in HzRGs (Cimatti \etal\ 1994).  
This suggestion is supported by the detection
of possibly extended, luminous X-ray emission surrounding some HzRGs 
(e.g. 3C356; see Crawford (1997) for a recent review).
The possibility that some HzRGs may be located at
the base of distant cooling flows would be 
an additional thread tying together the HzRGs and 
the low redshift cooling flows.  In addition
to supplying a scattering medium, cooling flows
are capable of fueling star formation and the central engine.

\section{Conclusions}

We have found a three-sigma upper limit 
to the degree of polarization of the $U$-band light emitted from A2597's blue
lobes to be less than $6\%$.  This limit includes
corrections for dilution by background starlight and nebular emission. 
The $U$-band emission from the blue lobes is composed
of $75-88\%$ stellar continuum from young stars, $13-18$\% 
nebular emission, and less than 6\% scattered light.
Earlier studies of the CDG in A1795 (McNamara \etal\ 1996a,b), 
which has similar properties to the CDG in A2597, 
came to similar conclusions.

Our limits do not support the conjecture that the blue
lobes are scattered light from an obscured BL Lac or
blazar nucleus associated with the FR I radio source PKS
2322$-$122. However, if the beamed AGN luminosity and
hence the polarized luminosity scales with radio luminosity,
the FR I radio galaxies in A2597 and A1795 could be scaled-down versions of the high redshift
radio galaxies exhibiting the alignment effect. AGN that would be 
present in these objects must have low $U$-band luminosities.
The asymmetries in the radio structures
of the A1795 and A2597 CDGs may have resulted from interactions between
the radio jets and dense, dusty clouds encountered by the jets.
The star formation associated with the blue lobes
may have been triggered by these interactions.

\acknowledgements

We thank Nicolas Cardiel for providing us with H$\beta$ equivalent widths.
B. R. M. was supported by
grant NAS8-39073 to the Smithsonian Astrophysical Observatory.
C. L. S. thanks Bill Sparks for a very useful conversation.
C. L. S. was supported in part by NASA Astrophysical Theory Program
grant 5-3057 and NASA ROSAT grants NAG 5-4787 and NAG 5-3308.

\newpage
\begin{center}
\begin{table*}
\caption{Polarization Measurements}
\vspace{2.5 mm}
\begin{tabular}{lccccccc}\hline\hline
Location& $r$ & PA & Ap &$P_{\rm tot}$  &$\sigma_{\rm p}$& $P_{\rm *}$ & $P_{\rm *,neb}$\\
~~& (arcsec) & (degrees) & (arcsec) &($\%$)&($\%$)&($\%$) &($\%$)\\
\hline
Nucleus& ... & ... &10.4  & $-0.9$  & 1.4 & $<4.9$&$<5.8$ \\
NE Lobe& 2.7& 45  & 4.70  & $~~0.5$ & 2.0 & $<4.8$&$<5.7$ \\
SW Lobe& 3.4& 236 & 4.70  & $-1.5$  & 2.0 & $<2.9$& $<3.4$\\
\hline\hline
\end{tabular}
\end{table*}
\end{center}
%\newpage

\begin{center}
\begin{table*}
\caption{Radio and Polarized Luminosities of Radio Galaxies}
\vspace{2.5 mm}
\begin{tabular}{lcccccc}\hline\hline
Object& $z$ &$S_{1.4}$ & $P_{1.4}$ &$P(U)$ &$L_{\rm pol}(U)$ &Refs\\
~~& ~~~ & (Jy) &(log[${\rm W~Hz}^{-1}$]) & ($\%$)&$(10^{42}~{\rm erg~sec}^{-1})$&~~~\\
\hline
3C 13 &1.351 & 5.56  &28.81 & 10 & 25& 1,7,10\\ 
3C 256&1.824 & 3.54  &28.90 &11 & 3.7& 2,11\\
3C 265&0.811 &12.98  &28.68 &10 & 8.9& 3,7,12\\
3C 324&1.206 &13.59  &29.08 &12 & 9.4& 4,11\\
3C 356&1.079 & 7.96  &28.75 &8  & 6.6& 1,7\\
3C 368&1.132 & 9.58  &28.87 &$<3~~~~~$  &$<34~~~~~~~~$ &5,7,11\\
3C 441&0.707 &11.55  &28.50 & 3  &2.2 &5,13\\
\hline
A2597  &0.082 &2.02 &25.79& $<6~~~~$& $<0.15~~~~$&8,9\\
A1795  &0.064 &0.97 &25.23& $<6~~~~$& $<0.04~~~~$&6,8\\
\hline\hline
\end{tabular}

1) Cimatti \etal\ (1997) 
2) Dey \etal\ (1996) \\
3) di Serego Alighieri \etal\ (1996) 
4) Cimatti \etal\ (1996)  \\
5) Dey (1998) 
6) McNamara \etal\ (1996) \\
7) White \& Becker (1992)
8) Heckman \etal\ (1989) \\
9) this paper 
10) Ficarra \etal\ (1985)\\
11) Wright \& Otrupcek (1990)
12) Becker \etal\ (1991)\\
13) Pilkington \& Scott (1965)
\end{table*}
\end{center}
\newpage
\clearpage

\centerline{\bf Figure Captions}
\noindent
{\bf Figure 1. } U-band composite grayscale image of the core of the A2597
 CDG, constructed as described in Section 3. The superposed white contours show
the 8.44 GHz radio continuum map. The blue lobes, shown in black, are
 coincident with the radio continuum. 
North is at top, east is to the left.

\noindent
{\bf Figure 2.} The U-band surface brightness profile (solid circles) in
magnitudes per square arcsec, with an arbitrary offset,
 plotted against semimajor axis to 
the $1/4$ power.  The solid line
represent the $R^{1/4}$-law surface brightness profile used to correct
the polarization measures for dilution by surrounding starlight.  
The blue lobes are located where the surface brightness profile rises
above the $R^{1/4}$ profile.

\noindent
{\bf Figure 3.} 
The predicted degree of polarization, $P$, if the blue lobes 
are due to scattered light from double beams of radiation.
The abscissa $\theta$ is the angle between the center of the cones of
radiation and our line-of-sight.
The predicted polarization includes the effect of the finite width of the
beams; it is averaged both along the line-of sight through each lobe and
across the projected width of the lobes.
The upper solid curve give the polarization if the lobes are due to
electron scattering (or Rayleigh scattering by small particles).
The lower short-dash curve is the predicted polarization if the lobes
are due to scattering by normal interstellar dust.
The lobes are asymmetric for dust scattering, and the curve gives the maximum
polarization of the two lobes.
The long-dash horizontal line gives the observed 3-$\sigma$ upper limit on
the polarization of 6\%.

\noindent
{\bf Figure 4.} 
Radio power is plotted against the polarized luminosity for alignment
effect radio galaxies.  The  3C radio
galaxies are grouped to the upper right, and the upper limits to
the polarized luminosity for A1795 and A2597 are in the lower left 
of the plot.  The solid line, normalized to the median of the 3C points,
represents $L(U)_{\rm pol}\propto P_{\rm rad}$.  The vertical dashed 
line indicates approximately the transition between FR I and FR II
radio luminosities.  

\clearpage
Figure 1. \\
\begin{figure}[h]
\hbox{
\hspace{1.3in}
\psfig{figure=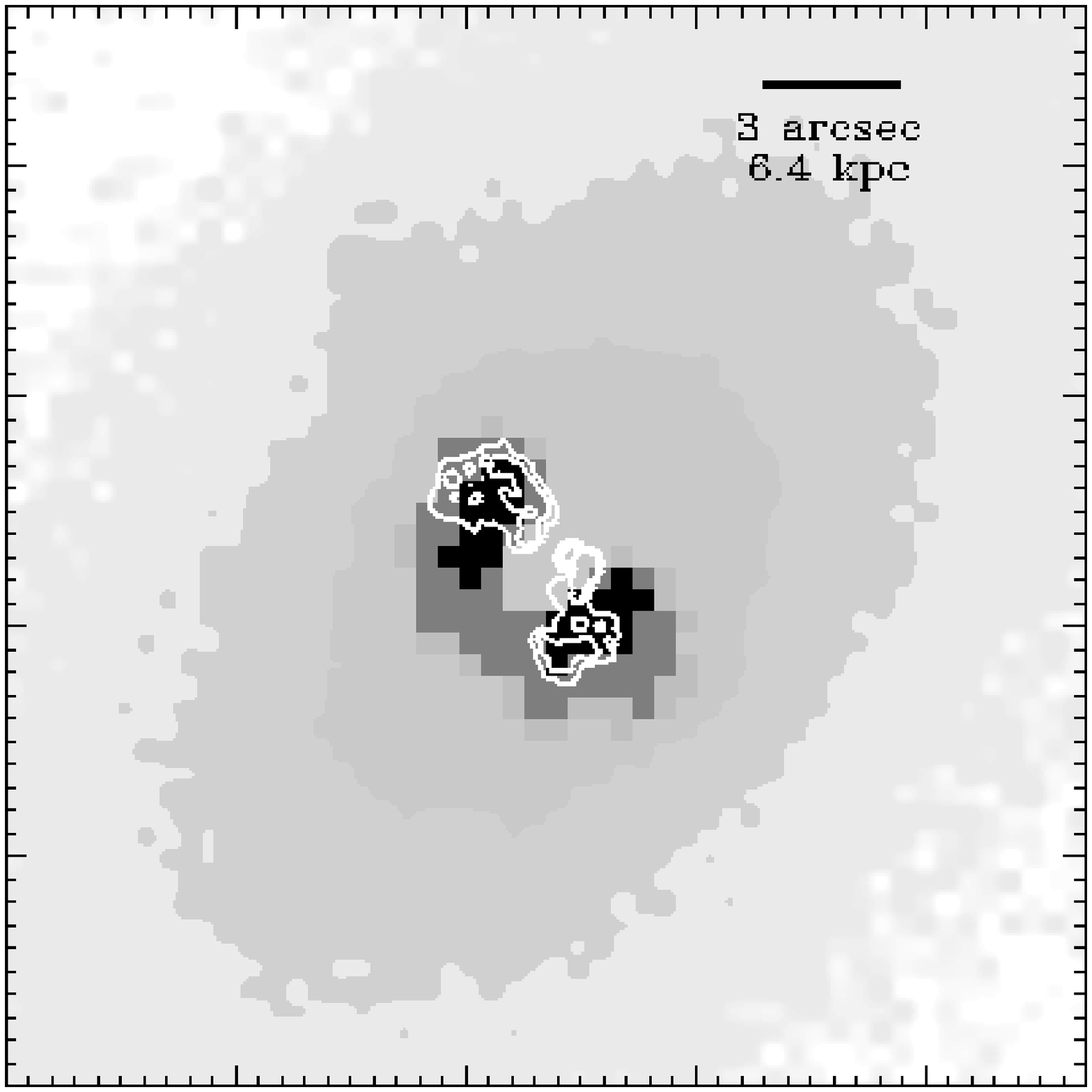,height=4in,width=4in}
}
\vspace{0.35in}
\end{figure}
\clearpage
Figure 2. \\
\begin{figure}[h]
\hbox{
\hspace{0in}
\psfig{figure=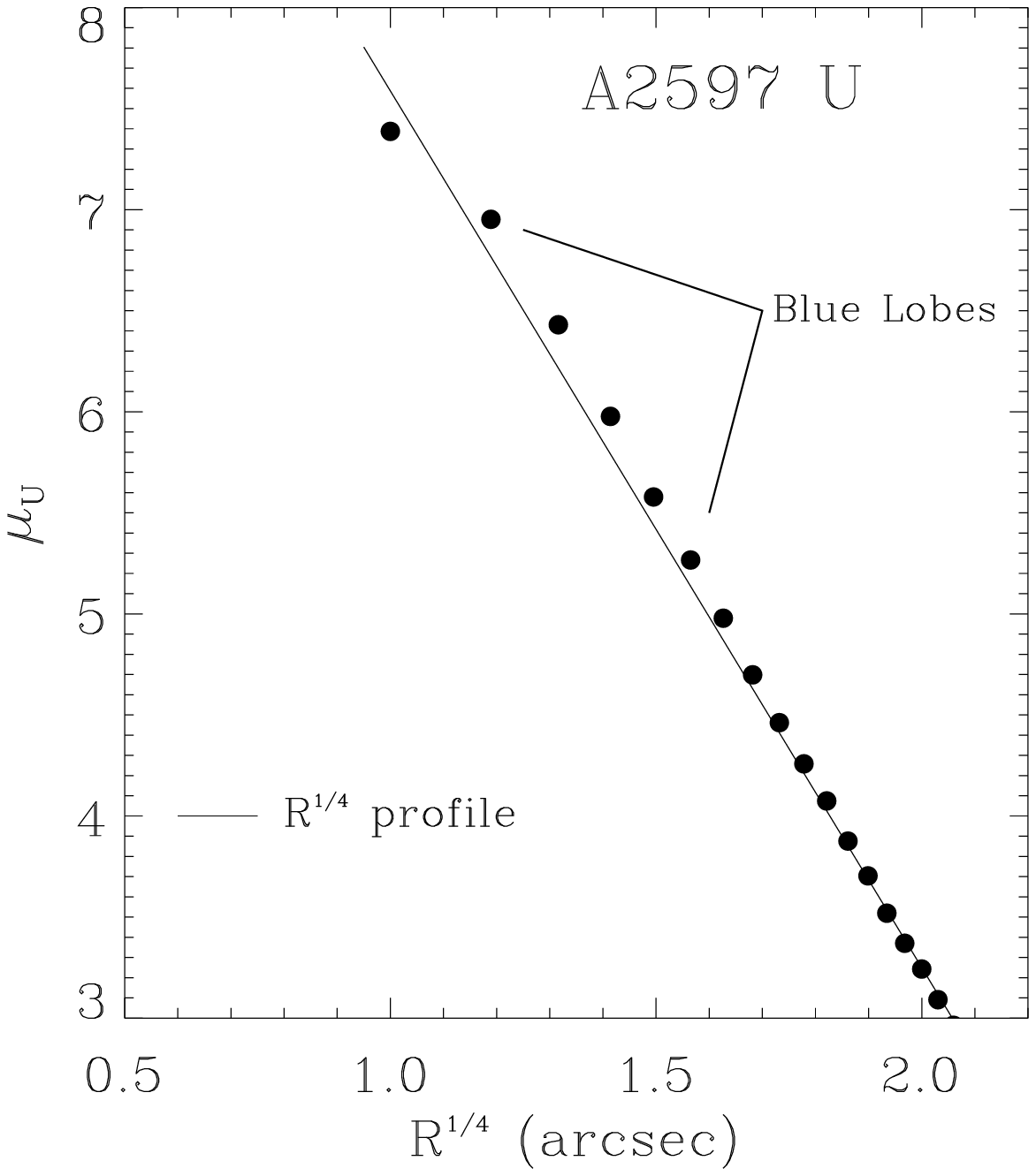,height=6in,width=6in}
}
\vspace{0.35in}
\end{figure}

%***************
\clearpage
\begin{figure}[t]
Figure 3. \\
\vskip 6.8truein
\includegraphics{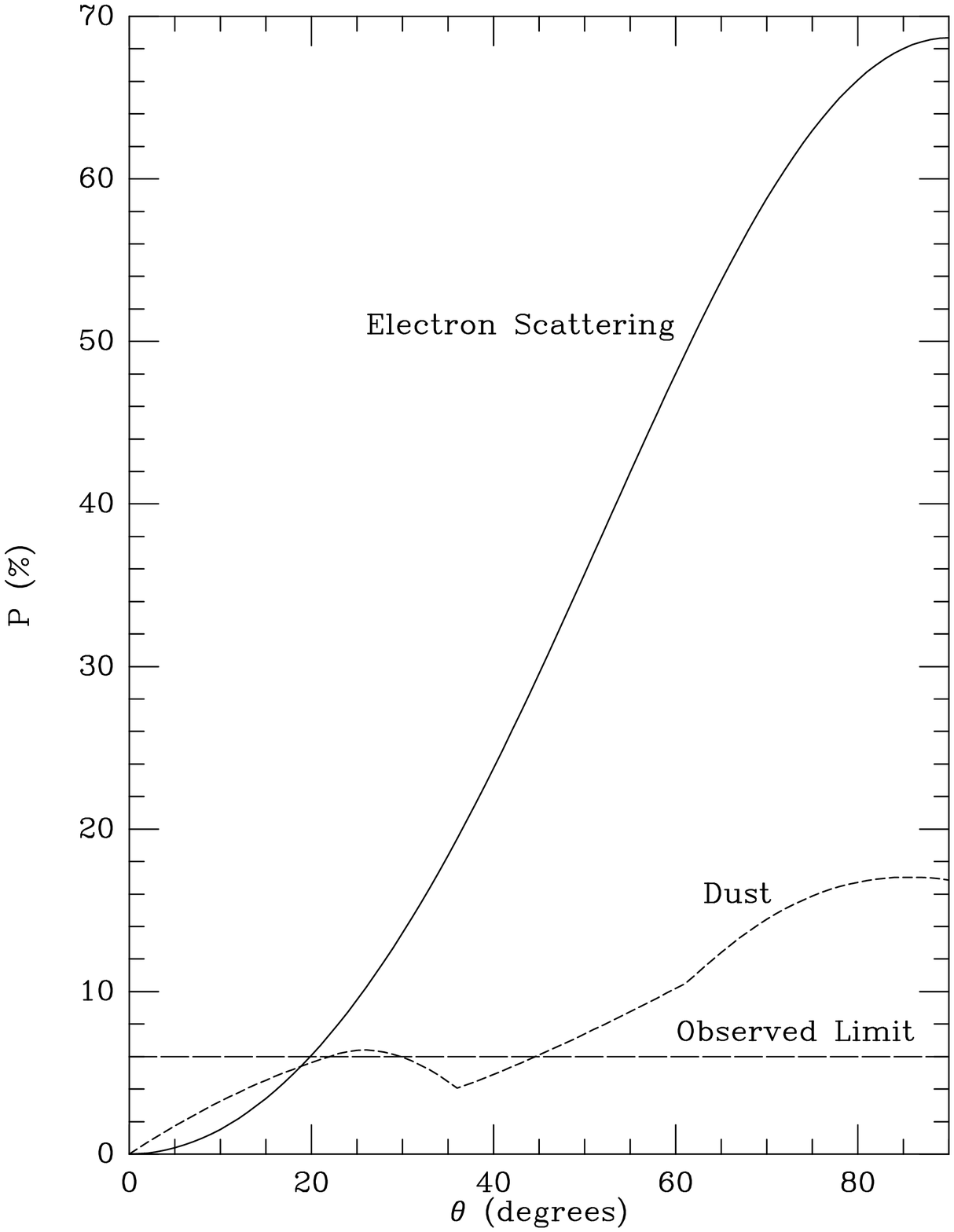}
\end{figure}
%******************
\clearpage
Figure 4. \\
\begin{figure}[h]
\hbox{
\hspace{1.3in}
\psfig{figure=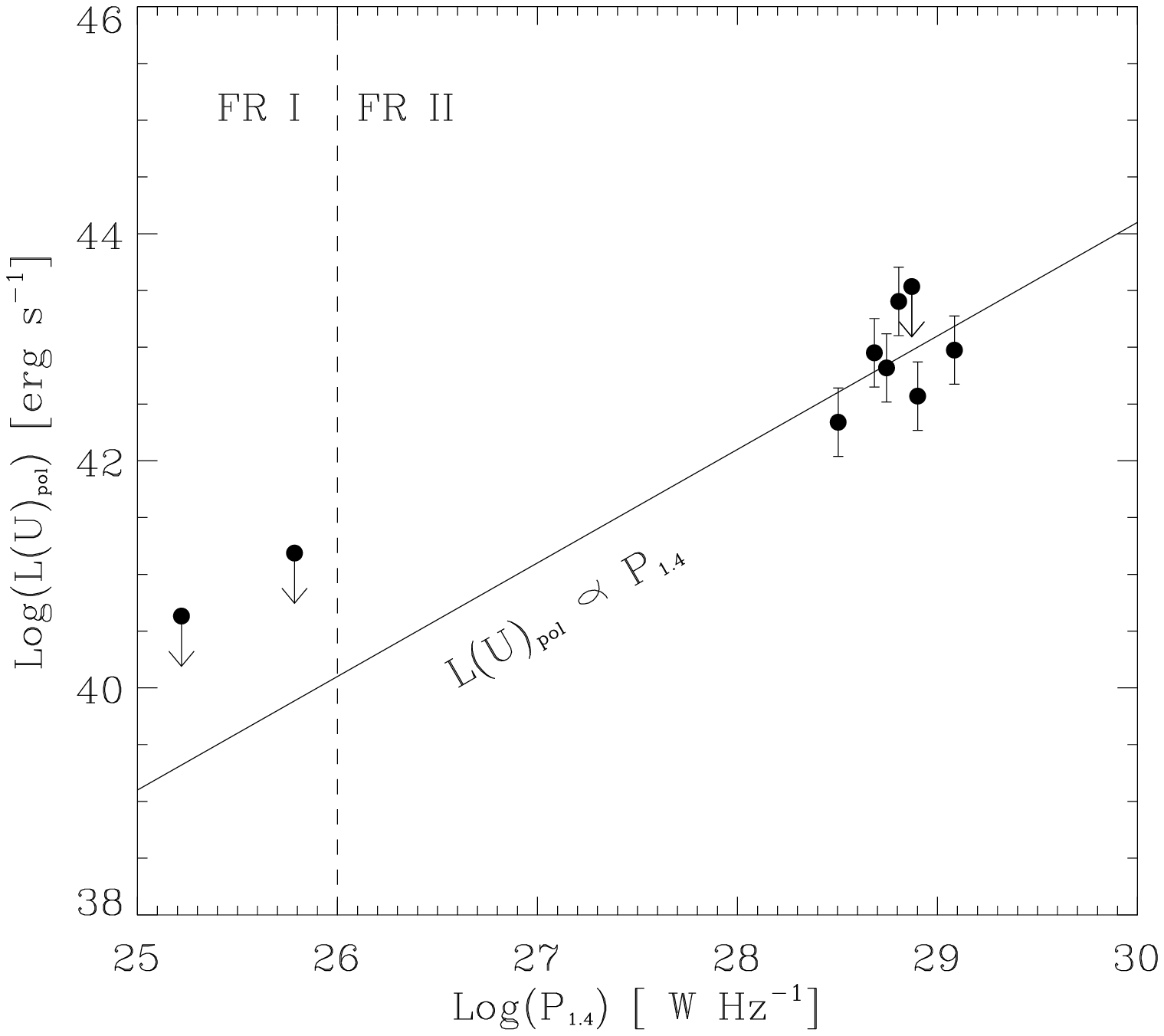,height=5in,width=5in}
}
\vspace{0.35in}
\end{figure}

% That's all, folks.
%
% The technique of segregating major semantic components of the document
% within "environments" is a very good one, but you as an author have to
% come up with a way of making sure each \begin{whatzit} has a corresponding
% \end{whatzit}.  If you miss one, LaTeX will probably complain a great
% deal during the composition of the document.  Occasionally, you get away
% with it right up to the \end{document}, in which case, you will see
% "\begin{whatzit} ended by \end{document}".

\end{document}

%% file: brmacro.tex
\def\etal{{\rm et~al\/}.}

\def\mathfont#1{\ifmmode{#1}\else{$#1$}\fi} %for math font     
\def\lae{\mathrel{<\kern-1.0em\lower0.9ex\hbox{$\sim$}}}  
\def\gae{\mathrel{>\kern-1.0em\lower0.9ex\hbox{$\sim$}}}  
\def\deg{\mathfont{^\circ}}  
\def\qo{\ifmmode{q_o}\else{$q_0$}\fi}  
\def\ho{\ifmmode{H_o}\else{$H_0$}\fi}   
\def\hounit{\ifmmode{{\rm km\  s}^{-1}\ {\rm Mpc}^{-        
1}}\else{${\rm    km\ s}^{-1}\ {\rm Mpc}^{-1}$}\fi}  
\def\zf{\ifmmode{z_{GF}}\else{$z_{GF}$}\fi}  
\def\kms{\ifmmode{{\rm km\ s}^{-1}}\else{${\rm km\ s}^{-1}$}\fi} 

\def\ergsec{\mathfont{ {\rm ergs\ s}^{-1}}}

\def\msun{\ifmmode{\ {\rm M}_\odot}\else{$ {\rm M}_\odot$}\fi}  
\def\msunyr{\ifmmode{\msun \ {\rm yr}^{-1}}\else{$\msun \ {\rm 
yr}^{-1}$}\fi}

\def\bv{\ifmmode{(B-V)}\else{$(B-V)$}\fi}
\def\sfr{\ifmmode{\dot S}\else{$\dot S$}\fi}

\def\ref{\noindent\hangindent=20truept\hangafter=1}

\def \apj            {ApJ} 
\def \apjlet        {ApJ}

\def \mnras          {MNRAS}

\def \vol#1          {${\bf \underline{#1}}$}

%% file: msbrm.bbl
\begin{references}
Aller, L.H. 1984, Physics of Thermal Gaseous Nebulae (Dordrecht: Reidel), 115, 98 \\
Antonucci, R. 1993, ARAA, 31, 473\\
Baum, S.A. 1992, in Clusters and Superclusters of Galaxies,
ed. A.C. Fabian, (Dordrecht: Kluwer), 171\\
Becker, R.H., White, R.L., Edwards, A.L. 1991, ApJS, 75, 1\\
Begelman, M.C., \& Cioffi, D.F. 1989, ApJ, 345, L21 \\
Bohren, C. F., \& Huffman, D. R. 1983, Absorption and Scattering of
Light by Small Particles (New York: Wiley)\\
Cardiel, N., Gorgas, J., Aragon-Salamanca, A. 1997, MNRAS, submitted\\
Cardiel, N., Gorgas, J., Aragon-Salamanca, A. 1998, in preparation.\\
Chambers, K.C., Miley, G.K., \& van Breugel, W. 1987, Nature, 329, 604\\
Cimatti, A., di Serego Alighieri, S. 1995, MNRAS, 273, L7\\
Cimatti, A., di Serego Alighieri, S., Field, G.B., Fosbury,
R.A.E. 1994, ApJ, 442, 562\\
Cimatti, A., Dey, A., van Breugel, W., Hurt, T., Antonucci,
R., 1997, ApJ, 476, 677\\
Cimatti, A., Dey, A., van Breugel, W., Antonucci, R., Spinrad, H. 1996, 465, 145\\
Crawford, C.S. 1997, in Galactic and Cluster Cooling Flows, ed. N. Soker, (San Francisco: Publ. Astr. Soc. Pacific), 38\\
Daly, R.A. 1990, \apj , 355, 416\\
Daly, R.A. 1992, \apj , 386, L9\\
Dey, A. 1998, to be published in ``The Most Distant Radio 
Galaxies'' \\
Dey, A., Cimatti, A., van Breugel, W., Antonucci, R., \& Spinrad, H. 1996, ApJ, 465, 157\\
Dey, A., van Breugel, W., Vacca, W.D., Antonucci, R. 1997, ApJ, 490, 698\\
De Young, D.S. 1989, \apjlet , 342, L59\\
De Young, D.S. 1995, ApJ, 446, 521\\
Dickson, R., Tadhunter, C., Shaw, M., Clark, N., Morganti, R., 1995, MNRAS, 273, L29\\
di Serego Alighieri, S., Fosbury, R.A.E., Quinn, P.J., \& Tadhunter, C.N. 1989, Nature, 341, 307\\
di Serego Alighieri, S., Cimatti, A., \& Fosbury, R.A.E. 1993, ApJ, 404, 584\\
di Serego Alighieri, S., Cimatti, A., \& Fosbury, R.A.E.,Perez-Fournon, I. 1996, MNRAS, 299, L57\\
Fabian, A.C. 1989, MNRAS, 238, 41p\\
Ficarra, A., Grueff, G., Tomassetti, G. 1985, AA Suppl, 59,255\\
Heckman, T.M., Baum, S.A., van Breugel W.J.M., \& McCarthy, P.J. 1989, \apj , 338, 48\\
Jannuzi, B.T. \& Elston, R. 1991, ApJ, 366, L69\\
Jannuzi, B. T., Elston, R., Schmidt, G. D., Smith, P. S., and Stockman, 
H. S. 1995, ApJ, 454, L111\\
Jannuzi, B. T. 1994, in I.A.U. Symposium No. 159, Multi-Wavelength Continuum Emission of AGN, ed. T. J. - L. Courvoisier \& A. Blecha, (Dordrecht; Kluwer), 470\\
McCarthy, P.J. 1993, ARAA, 31, 639\\
McNamara, B.R., \& O'Connell, R.W. 1993, AJ, 105, 417 (MO93)\\
McNamara, B.R., Jannuzi, B.T., Sarazin, C.L., Elston, R., \& Wise, M. 1996a, ApJ, 469,66 \\
McNamara, B.R., Wise, M., Sarazin, C.L., Jannuzi, B.T., \& Elston, R. 1996b, ApJ, 466, L9\\
McNamara, B.R., O'Connell, R.W., \& Sarazin, C.L. 1996c, AJ, 112, 91\\
McNamara, B.R., 1997, in Galactic and Cluster Cooling Flows,
ed. N. Soker, (San Francisco: Publ. Astr. Soc. Pacific), 109\\
Murphy, B.W., \& Chernoff, D.F., 1993, ApJ, 418, 60\\
O'Dea, C.P., Gallimore, J.F., \& Baum, S.A. 1995, AJ, 109, 26\\
Osterbrock, D.E. 1974, Astrophysics of Gaseous Nebulae, (San Francisco: Freeman), 69\\
Padovani, P. \& Urry, C.M. 1990, ApJ, 356, 75\\
Pilkington, J.D., Scott, P.F. 1965, MNRAS, 69, 183\\
Pinkney, J., Holtzman, J.A., Garasi, C., et al. 1996, ApJ, 468, L13\\
Rees, M.J. 1989, \mnras , 239, 1p\\
Rigler, M.A., Lilly, S.J., Stockton, A., Hammer, F., \& Le Fevre, O. 1992, ApJ, 385, 61\\
Sarazin, C.L., \& Wise, M.W. 1993, ApJ, 411, 55\\
Sarazin, C.L., Burns, J.O., Roettiger, K., \& McNamara,
B.R. 1995, ApJ, 447, 559 \\ 
Scarrott, S.M., Rolph, C.D., \& Tadhunter, C.N. 1990, MNRAS, 243, 5p \\
Urry, C.M., Padovani, P. 1995, PASP, 107, 803\\
van Breugel, W, Heckman, T., \& Miley, G. 1984, ApJ, 276, 79\\
van Breugel, W., Filippenko, A.V., Heckman, T., \& Miley, G. 1985, ApJ, 293, 83\\
van Breugel, W., \& Dey, A. 1993, ApJ, 414, 563\\
Voit, G.M., \& Donahue, M. 1997, ApJ, 486, 242\\
White, R. L. 1979, ApJ, 229, 954\\
White, R. L., Becker, R.H. 1992, ApJS, 79, 331\\
Wright, A. Otrupcek, R. 1990, Parkes Catalog, p \\
Zellner, B. 1973, in I.A.U. Symp.\ 52: Interstellar Dust and Related Topics, ed.\ J. M Greenberg \& H. C. Van de Hulst (Dordrecht: Reidel), 109\\
\end{references}
